\documentclass[12pt,twoside,a4paper]{article} 
\usepackage{natbib}
\usepackage{bled2e}
\usepackage{amsmath}
\usepackage{bm}
\usepackage[justification=centering,margin=10pt,font=small,labelfont=bf]{caption}

\usepackage{graphicx}
\usepackage{array}

\usepackage{times}

\makeatletter
\@addtoreset{equation}{section}
\makeatother
\title{Estimating Bayes factors from minimal summary statistics in repeated measures analysis of variance designs}
\author{
  Thomas J. Faulkenberry\thanks{~~Department of Psychological Sciences, Tarleton State University; faulkenberry@tarleton.edu}
}

\date{ }
\begin{document}

\maketitle

\markboth{Thomas J. Faulkenberry}
         {Bayes factors for ANOVA summaries\ldots}

\begin{abstract}
\setcounter{footnote}{1}
  In this paper, I develop a formula for estimating Bayes factors directly from minimal summary statistics produced in repeated measures analysis of variance designs. The formula, which requires knowing only the $F$-statistic, the number of subjects, and the number of repeated measurements per subject, is based on the BIC approximation of the Bayes factor, a common default method for Bayesian computation with linear models. In addition to providing computational examples, I report a simulation study in which I demonstrate that the formula compares favorably to a recently developed, more complex method that accounts for correlation between repeated measurements. The minimal BIC method provides a simple way for researchers to estimate Bayes factors from a minimal set of summary statistics, giving users a powerful index for estimating the evidential value of not only their own data, but also the data reported in published studies.
\end{abstract}


\section{Introduction}

In this paper, I discuss how to apply the BIC approximation \citep{kassRaftery1995,wagenmakers2007,masson2011,nathoo2016} to compute Bayes factors for repeated measures experiments using only minimal summary statistics from the analysis of variance \citep[e.g.,][]{ly2018,faulkenberry2018}. Critically, I develop a formula (Equation \ref{eq:bic2}) that works for repeated measures experiments. Further, I investigate its performance against a method of \citet{nathoo2016} which accounts for varying levels of correlation between repeated measurements. Among several ``default prior'' solutions to computing Bayes factors for common experimental designs \citep{rouder2009,rouder2012}, each of which requires raw data for computation, the proposed formula stands out for providing the user with a simple expression for the Bayes factor that can be computed even when only the summary statistics are known. Thus, equipped with only a hand calculator, one can immediately estimate a Bayes factor for many results reported in published paper (even null effects), providing a meta-analytic tool that can be quite useful when trying to establish the evidential value of a collection of published results.

\section{Background}
To begin, let us consider the elementary case of a one-factor independent groups design. Consider a set of data $y_{ij}$, on which we impose the linear model%
\[
  y_{ij}=\mu + \alpha_j + \varepsilon_{ij}; \hspace{3mm} i=1,\cdots,n; \hspace{1mm} j=1,\dots,k
\]%
where $\mu$ represents the grand mean, $\alpha_j$ represents the treatment effect associated with group $j$, and $\varepsilon_{ij} \sim \mathcal{N}(0,\sigma_{\varepsilon}^2)$. In all, we have $N=nk$ independent observations. To proceed with hypothesis testing, we define two competing models:%
\begin{gather*}
  \mathcal{H}_0:\alpha_j=0\text{ for }j=1,\dots,k\\
  \mathcal{H}_1:\alpha_j\neq 0\text{ for some }j
\end{gather*}

Classically, model selection is performed using the analysis of variance (ANOVA), introduced in the 1920s by Sir Ronald Fisher \citep{fisher1925}. Roughly, ANOVA works by partitioning the total variance in the data $\bm{y}$ into two sources -- the variance between the treatment groups, and the residual variance that is left over after accounting for this treatment variability. Then, one calculates an $F$ statistic, defined as the ratio of the between-groups variance to the residual variance. Inference is then performed by quantifying the likelihood of the observed data $\bm{y}$ under the null hypothesis $\mathcal{H}_0$. Specifically, this is done by computing the probability of obtaining the observed $F$ statistic (or greater) under $\mathcal{H}_0$. If this probability, called the $p$-value, is small, this indicates that the data $\bm{y}$ are {\it rare} under $\mathcal{H}_0$, so the researcher may reject $\mathcal{H}_0$ in favor of the alternative hypothesis $\mathcal{H}_1$. Though it is a classic procedure, some issues arise that make it problematic. First, the $p$-value is not equivalent to the posterior probability $p(\mathcal{H}_0\mid \bm{y})$. Despite this distinction, many researchers incorrectly believe that a $p$-value directly indexes the probability that $\mathcal{H}_0$ is true \citep{gigerenzer2004}, and thus take a small $p$-value to represent evidence for $\mathcal{H}_1$. However, \citet{berger1987} demonstrated that $p$-values classically overestimate this evidence. For example, with a $t$-test performed on a sample size of 100, a $p$-value of 0.05 transforms to $p(\mathcal{H}_0\mid \bm{y})=0.52$ -- rather than reflecting evidence for $\mathcal{H}_1$, this small $p$-value reflects data that slightly prefers $\mathcal{H}_0$. Second, the ``evidence'' provided for $\mathcal{H}_1$ via the $p$-value is only indirect, as the $p$-value only measures the predictive adequacy of $\mathcal{H}_0$; the $p$-value procedure makes no such measurement of predictive adequacy for $\mathcal{H}_1$.

For these reasons, I will consider a Bayesian approach to the problem of model selection. The approach I will describe in this paper is to compute the Bayes factor \citep{kassRaftery1995}, denoted $\text{BF}_{01}$, for $\mathcal{H}_0$ over $\mathcal{H}_1$. In general, the Bayes factor is defined as the ratio of marginal likelihoods for $\mathcal{H}_0$ and $\mathcal{H}_1$, respectively. That is,%
\begin{equation}\label{eq:marginal}
 \text{BF}_{01} = \frac{p(\bm{y}\mid \mathcal{H}_0)}{p(\bm{y}\mid \mathcal{H}_1)}.
\end{equation}%
This ratio is immediately useful in two ways. First, it indexes the relative likelihood of observing data $\bm{y}$ under $\mathcal{H}_0$ compared to $\mathcal{H}_1$, so $\text{BF}_{01}>1$ is taken as evidence for $\mathcal{H}_0$ over $\mathcal{H}_1$. Similarly, $\text{BF}_{01}<1$ is taken as evidence for $\mathcal{H}_1$. Second, the Bayes factor indicates the extent to which the prior odds for $\mathcal{H}_0$ over $\mathcal{H}_1$ are updated after observing data. Said differently, the ratio of posterior probabilities for $\mathcal{H}_0$ and $\mathcal{H}_1$ can be found by multiplying the ratio of prior probabilities by $\text{BF}_{01}$ (a fact which follows easily from Bayes' theorem):%
\begin{equation}\label{eq:bayesUpdating}
  \frac{p(\mathcal{H}_0\mid \bm{y})}{p(\mathcal{H}_1\mid \bm{y})} =  \text{BF}_{01}\cdot \frac{p(\mathcal{H}_0)}{p(\mathcal{H}_1)}.
\end{equation}%
One interesting consequence of Equation \ref{eq:bayesUpdating} is that we can use the Bayes factor to compute the posterior probability of $\mathcal{H}_0$ as a function of the prior model probabilities. To see this, consider the following. If we solve Equation \ref{eq:bayesUpdating} for the posterior probability $p(\mathcal{H}_0\mid \bm{y})$ and then use Bayes' theorem, we see%
\begin{align*}
  p(\mathcal{H}_0\mid \bm{y}) &= \text{BF}_{01} \cdot \frac{p(\mathcal{H}_0)}{p(\mathcal{H}_1)} \cdot p(\mathcal{H}_1\mid \bm{y})\\
                              &= \frac{\text{BF}_{01} \cdot p(\mathcal{H}_0) \cdot p(\bm{y}\mid \mathcal{H}_1)\cdot p(\mathcal{H}_1)}{p(\mathcal{H}_1)\cdot p(\bm{y})}\\
  &= \frac{\text{BF}_{01}\cdot p(\mathcal{H}_0)\cdot p(\bm{y}\mid \mathcal{H}_1)}{p(\bm{y}\mid \mathcal{H}_0)\cdot p(\mathcal{H}_0) + p(\bm{y}\mid \mathcal{H}_1)\cdot p(\mathcal{H}_1)}.
\end{align*}%
Dividing both numerator and denominator by the marginal likelihood $p(\bm{y}\mid \mathcal{H}_1)$ gives us%
\[
  p(\mathcal{H}_0\mid \bm{y}) = \frac{\text{BF}_{01}\cdot p(\mathcal{H}_0)}{\text{BF}_{01}\cdot p(\mathcal{H}_0) + p(\mathcal{H}_1)}.
\]%
By Equation \ref{eq:marginal}, we have $\text{BF}_{10} = 1/\text{BF}_{01}$. It can then be shown similarly that
\[
  p(\mathcal{H}_1\mid \bm{y}) = \frac{\text{BF}_{10}\cdot p(\mathcal{H}_1)}{\text{BF}_{10}\cdot p(\mathcal{H}_1) + p(\mathcal{H}_0)}.
\]%
In practice, researchers often assume both models are {\it a priori} equally likely, and thus set both $p(\mathcal{H}_0)=p(\mathcal{H}_1) = 0.5$. In this case, we obtain the simplified forms%
\begin{equation}\label{eq:probability}
  p(\mathcal{H}_0\mid \bm{y}) = \frac{\text{BF}_{01}}{\text{BF}_{01}+1}, \hspace{1cm} p(\mathcal{H}_1\mid \bm{y}) = \frac{\text{BF}_{10}}{\text{BF}_{10}+1}.
\end{equation}

Though there are many simple quantities that can be derived from the Bayes factor, the actual computation of $\text{BF}_{01}$ can be quite difficult, as the marginal likelihoods in Equation \ref{eq:marginal} each require integrating over a prior distribution of model parameters. This often results in integrals that do not admit closed form solutions, requiring approximate techniques to estimate the Bayes factor. In \citet{faulkenberry2018}, it was shown that for an independent groups design, one can use the $F$-ratio and degrees of freedom from an analysis of variance to compute an approximation of $\text{BF}_{01}$ that is based on a unit information prior \citep{wagenmakers2007,masson2011}. Specifically%
\begin{equation}\label{eq:bic}
  \text{BF}_{01} \approx \sqrt{N^{df_1}\Bigl(1+\frac{Fdf_1}{df_2}\Bigr)^{-N}},
\end{equation}%
where $F(df_1,df_2)$ is the $F$-ratio from a standard analysis of variance applied to these data.

As an example, consider a hypothetical dataset containing $k=4$ groups of $n=25$ observations each (for a total of $N=100$ independent observations). Suppose that an ANOVA produces $F(3,96)=2.76$, $p=0.046$. This result would be considered as ``statistically significant'' by conventional null hypothesis standards, and traditional practice would dictate that we reject $\mathcal{H}_0$ in favor of $\mathcal{H}_1$. But is this result really evidential for $\mathcal{H}_1$? Applying Equation \ref{eq:bic} shows:%
\begin{align*}
  \text{BF}_{01} & \approx \sqrt{N^{df_1}\Bigl(1+\frac{Fdf_1}{df_2}\Bigr)^{-N}}\\
         & = \sqrt{100^3\Bigl(1+\frac{0.76\cdot 3}{96}\Bigr)^{-100}}\\
  &= 15.98.
\end{align*}%
This result indicates quite the opposite: by definition of the Bayes factor, this implies that the observed data are almost 16 times more likely under $\mathcal{H}_0$ than $\mathcal{H}_1$. Note that the appearance of such contradictory conclusions from two different testing frameworks is actually a classic result known as Lindley's paradox \citep{lindley1957}.

\section{The BIC approximation for repeated measures}

Against this background, the goal now is to extend Equation \ref{eq:bic} to the case where we have an experimental design with repeated measurements. For context, consider an experiment where $k$ measurements are taken from each of $n$ experimental subjects. We then have a total of $N=nk$ observations, but they are no longer independent measurements. Assume a linear mixed model structure on the observations:%
\[
  y_{ij} = \mu + \alpha_j + \pi_i + \varepsilon_{ij}; \hspace{3mm} i=1,\dots,n;\hspace{1mm}j=1\cdots,k,
\]%
where $\mu$ represents the grand mean, $\alpha_j$ represents the treatment effect associated with group $j$, $\pi_i$ represents the effect of subject $i$, and $\varepsilon_{ij} \sim \mathcal{N}(0,\sigma_{\varepsilon}^2)$. Due to the correlated structure of these data, we have $n(k-1)$ independent observations.  We will define models $\mathcal{H}_0$ and $\mathcal{H}_1$ as above. Also, we will denote the sums of squares terms in the model in the usual way, where%
\[
  SSA = n\sum_{j=1}^k(\overline{y}_{\cdot j}-\overline{y}_{\cdot \cdot})^2, \hspace{1cm}SSB = k\sum_{i=1}^n(\overline{y}_{i\cdot}-\overline{y}_{\cdot \cdot})^2
\]%
represent the sums of squares corresponding to the treatment effect and the subject effect, respectively,%
\[
  SST = \sum_{i=1}^n \sum_{j=1}^k (y_{ij} - \overline{y}_{\cdot \cdot})^2
\]%
represents the total sum of squares, and%
\[
  SSR = SST - SSA - SSB
\]%
represents the residual sum of squares left over after accounting for both treatment and subject effects. From here, we can compute the $F$-statistic for the treatment effect in our design as%
\[
  F=\frac{SSA}{SSR}\cdot \frac{df_{\text{residual}}}{df_{\text{treatment}}} = \frac{SSA}{SSR}\cdot \frac{(n-1)(k-1)}{k-1} = \frac{SSA}{SSR}\cdot (n-1).
    \]%
We will now show that this $F$ statistic can be used to estimate $\text{BF}_{01}$. 
    
To this end, note the following. Prior work of \citet{wagenmakers2007} has shown that $\text{BF}_{01}$ can be approximated as%
\[
  \text{BF}_{01} \approx \exp(\Delta BIC_{10}/2),
\]%
where%
\[
  \Delta BIC_{10} = N\ln \Biggl(\frac{SSE_1}{SSE_0}\Biggr) + (\kappa_1-\kappa_0)\ln(N).
\]%
Here, $N$ is equal to the number of independent observations; as noted above, this is equal to $n(k-1)$ for our repeated measures design. $SSE_1$ represents the variability left unexplained by $\mathcal{H}_1$; for our design, this is equal to the residual sum of squares, $SSR$. $SSE_0$ represents the variability left unexplained by $\mathcal{H}_0$; for our design, this is equal to the sum of the treatment sum of squares and the residual sum of squares, $SSA+SSR$. Finally, $\kappa_1-\kappa_0$ is equal to the difference in the number of parameters between $\mathcal{H}_1$ and $\mathcal{H}_0$; this is equal to $k-1$ for our design. 

We are now ready to derive a formula for $\text{BF}_{01}$. First, we will re-express $\Delta BIC_{10}$ in terms of $F$:%
\begin{align*}
  \Delta BIC_{10} &= N\ln \Biggl(\frac{SSE_1}{SSE_0}\Biggr) + (\kappa_1-\kappa_0)\ln(N)\\
                  &= n(k-1)\ln\Biggl(\frac{SSR}{SSR+SSA}\Biggr) + (k-1)\ln\Bigl(n(k-1)\Bigr)\\
                  &= n(k-1)\ln\Biggl(\frac{1}{1+\frac{SSA}{SSR}}\Biggr) + (k-1)\ln\Bigl(n(k-1)\Bigr)\\
                  &= n(k-1)\ln\Biggl(\frac{n-1}{n-1+\frac{SSA}{SSR}\cdot (n-1)}\Biggr) + (k-1)\ln\Bigl(n(k-1)\Bigr)\\
  &= n(k-1)\ln\Biggl(\frac{n-1}{n-1+F}\Biggr) + (k-1)\ln\Bigl(n(k-1)\Bigr)\\
\end{align*}%
Thus, we can write%
\begin{align*}
  \text{BF}_{01} &\approx \exp(\Delta BIC_{10}/2)\\
          &= \exp \Biggl[ \frac{n(k-1)}{2}\ln \Biggl(\frac{n-1}{n-1+F}\Biggr) + \frac{k-1}{2}\ln\Bigl(n(k-1)\Bigr)\Biggr]\\
          &= \Biggl(\frac{n-1}{n-1+F}\Biggr)^{\frac{n(k-1)}{2}}\cdot \Bigl(n(k-1)\Bigr)^{\frac{k-1}{2}}\\
          &= \sqrt{\Bigl(n(k-1)\Bigr)^{k-1}\cdot \Biggl(\frac{n-1}{n-1+F}\Biggr)^{n(k-1)}}\\
          &= \sqrt{(nk-n)^{k-1}\cdot \Biggl(\frac{n-1}{n-1+F}\Biggr)^{nk-n}}
\end{align*}%
If we invert the term containing $F$ and divide $n-1$ into the resulting numerator, we get the following formula:%
\begin{equation}\label{eq:bic2}
          \text{BF}_{01} \approx \sqrt{(nk-n)^{k-1}\cdot \Biggl(1+\frac{F}{n-1}\Biggr)^{n-nk}},
\end{equation}%
where $n$ equals the number of subjects and $k$ equals the number of repeated measurements per subject.

I will now give an example of using Equation \ref{eq:bic2} to compute a Bayes factor. The example below is based on data from \citet{faulkenberryBowman2018}. In this experiment, subjects were presented with pairs of single digit numerals and asked to choose the numeral that was presented in the larger font size. For each of $n=23$ subjects, response times were recorded in $k=2$ conditions -- congruent trials and incongruent trials. Congruent trials were defined as those in which the physically larger digit was also the numerically larger digit (e.g., {\scriptsize 2} -- {\large 8}). Incongruent trials were defined such that the physically larger digit was numerically smaller (e.g., {\large 2} -- {\scriptsize 8}). \citet{faulkenberryBowman2018} then fit each subjects' {\it distribution} of response times to a parametric model \citep[a shifted Wald model; see][for details]{anders2016,faulkenberry2017}, allowing them to investigate the effects of congruity on shape, scale, and location of the response time distributions. Specifically, they predicted that the leading edge, or {\it shift}, of the distributions would not differ between congruent and incongruent trials, thus providing support against an early encoding-based explanation of the observed size-congruity effect \citep{santens2011,faulkenberry2016,sobel2016,sobel2017}. The shift parameter was calculated for both of the $k=2$ congruity conditions for each of the $n=23$ subjects. The resulting ANOVA summary table is presented in Table \ref{tab:anova2}.

\begin{table}[h!]
  \centering
  \begin{tabular}%
    {p{2cm}%
    >{\raggedleft\arraybackslash}p{2cm}%
    >{\raggedleft\arraybackslash}p{1cm}%
    >{\raggedleft\arraybackslash}p{2cm}%
    >{\raggedleft\arraybackslash}p{2cm}%
    >{\raggedleft\arraybackslash}p{2cm}%
    }
    Source & $SS$ & $df$ & $MS$ & $F$ & $p$\\
    \hline
    Subjects & 103984 & 22 & 4727 & &\\
    Treatment & 739 & 1 & 739 & 1.336 & $0.26$\\
    Residual & 12176 & 22 & 553 & & \\
    Total & 116399 & 45& & & \\
    \hline
  \end{tabular}             
  \caption{ANOVA summary table for shift parameter data of \citet{faulkenberryBowman2018}}
  \label{tab:anova2}
\end{table}

Applying the minimal BIC method from Equation \ref{eq:bic} gives us the following:%
\begin{align*}
  \text{BF}_{01} &\approx \sqrt{(nk-n)^{k-1}\cdot \Biggl(1+\frac{F}{n-1}\Biggr)^{n-nk}}\\
         &= \sqrt{(23\cdot 2-23)^{2-1}\Biggl(1+\frac{1.336}{23-1}\Biggr)^{(23-23\cdot 2)}}\\
         &= \sqrt{23^1\Biggr(1+\frac{1.336}{22}\Biggr)^{-23}}\\
  &=2.435
\end{align*}%
This Bayes factor tells us that the observed data are approximately 2.4 times more likely under $\mathcal{H}_0$ than $\mathcal{H}_1$. Assuming equal prior model odds, we use Equation \ref{eq:probability} to convert the Bayes factor to a posterior model probability, giving positive evidence for $\mathcal{H}_0$:%
\begin{align*}
  p(\mathcal{H}_0\mid \bm{y}) &= \frac{\text{BF}_{01}}{\text{BF}_{01}+1}\\
                                   &= \frac{2.435}{2.435+1}\\
  &= 0.709.
\end{align*}

\section{Accounting for correlation between repeated measurements}

In a recent paper, \citet{nathoo2016} took a slightly different approach to calculating Bayes factors for repeated measures designs, investigating the role of {\it effective sample size} in repeated measures designs \citep{jones2011}. For single-factor repeated measures designs, effective sample size is defined as%
\[
  n_{\text{eff}}=\frac{nk}{1+\rho(k-1)},
\]%
where $\rho$ is the intraclass correlation,%
\[
  \rho=\frac{\sigma_{\pi}^2}{\sigma_{\pi}^2+\sigma_{\varepsilon}^2}.
\]%
Thus, $\rho=0$ implies $n_{\text{eff}}=nk$, whereas $\rho=1$ implies $n_{\text{eff}}=n$. Though $\rho$ is unknown, \citet{nathoo2016} developed a method to estimate it from $SS$ values in the ANOVA, leading to the following:%
\begin{align*}
  \Delta BIC_{10} &= n(k-1)\ln\Biggl(\frac{SST-SSA-SSB}{SST-SSB}\Biggr)\\ 
                  &+(k+2)\ln\Biggl(\frac{n(SST-SSA)}{SSB}\Biggr)\\
  &-3\ln\Biggl(\frac{nSST}{SSB}\Biggr) 
\end{align*}%
This estimate provides a better account of the correlation between repeated measurements, but the benefit comes at a price of added complexity, and it is not clear how to reduce this formula to a simple expression involving only $F$ as we do with Equation \ref{eq:bic2}. This leads to the natural question: how well does the minimal BIC method from Equation \ref{eq:bic2} match up with the more complex approach of \citet{nathoo2016}?

As a first step toward answering this question, let us revisit the example presented above. We can apply the Nathoo and Masson formula to the ANOVA summary in Table \ref{tab:anova2}:%
\begin{align*}
  \Delta BIC_{10} & = 23(2-1)\ln\Biggl(\frac{116399-739-103984}{116399-103984}\Biggr)\\
                  & \hspace{5mm} + (2+2)\ln\Biggl(\frac{23(116399-739)}{103984}\Biggr)\\
                  & \hspace{5mm} -3\ln\Biggl(\frac{23(116399)}{103984}\Biggr)\\
                  &= 23\ln(0.9405) + 4\ln(25.583) - 3\ln(25.746)\\
  &= 1.812.
\end{align*}%
This equates to a Bayes factor of%
\begin{align*}
  \text{BF}_{01} & =\exp(\Delta BIC_{10}/2)\\
         & = \exp(1.812/2) \\
         & = 2.474
\end{align*}%
and a posterior model probability of $p(\mathcal{H}_0\mid \bm{y})=2.474/(2.474+1)= 0.712$.  Clearly, these computations are quite similar to the ones we performed with Equation \ref{eq:bic2}, with both methods indicating positive evidence for $\mathcal{H}_0$ over $\mathcal{H}_1$.

\section{Simulation study}

The computations in the previous section reflect two preliminary facts. First, the method of \citet{nathoo2016} yields Bayes factors and posterior model probabilities that take into account an estimate of the correlation between repeated measurements. This is a highly principled approach which the minimal BIC method of Equation \ref{eq:bic2} does not take. However, as we can see with both computations, the general conclusion remains the same regardless of whether we use the minimal BIC method or the method of Nathoo and Masson. Given that our Equation \ref{eq:bic2} is (1) easy to use, and (2) requires only three inputs (the number of subjects $n$, the number of repeated measurement conditions $k$, and the $F$ statistic), we wonder if the minimal BIC method produces results that are sufficient for day-to-day work, with the risk of being conservative being outweighed by its simplicity. To answer this question, I conducted a Monte Carlo simulation\footnote{The simulation script (in R) and resulting simulated datasets can be downloaded from https://git.io/Jfekh.} to systematically investigate the relationship between Equation \ref{eq:bic2} and the Nathoo and Masson method across a wide variety of randomly generated datasets. 

In this simulation, I randomly generated datasets that reflected the repeated measures designs that we have discussed throughout this paper. Specifically, data were generated from the linear mixed model%
\[
  Y_{ij} = \mu + \alpha_j + \pi_i + \varepsilon_{ij};\hspace{5mm}i=1,\dots,n;\hspace{3mm}j=1,\dots,k,
\]%
where $\mu$ represents a grand mean, $\alpha_j$ represents a treatment effect, and $\pi_i$ represents a subject effect. For convenience, I set $k=3$, though similar results were obtained with other values of $k$ (not reported here). Also, I assumed $\pi_i\sim \mathcal{N}(0,\sigma_{\pi}^2)$ and $\varepsilon_{ij}\sim \mathcal{N}(0,\sigma_{\varepsilon}^2)$. I then systematically varied three components of the model:

\begin{enumerate}
\item The number of subjects $n$ was set to either $n=20$, $n=50$, or $n=80$;
\item The intraclass correlation $\rho$ between treatment conditions was set to be either $\rho=0.2$ or $\rho=0.8$; 
\item The size of the treatment effect was manipulated to be either null, small, or medium. Specifically, these effects were defined as follows. Let $\mu_j = \mu+\alpha_j$ (i.e., the condition mean for treatment $j$). Then we define effect size as%
\[
  \delta = \frac{\max(\mu_j)-\min(\mu_j)}{\sqrt{\sigma_{\pi}^2 + \sigma_{\varepsilon}^2}},
\]%
and correspondingly, we set $\delta$ to one of three values: $\delta=0$ (null effect), $\delta=0.2$ (small effect), and $\delta=0.5$ (medium effect). Also note that since we can write the intraclass correlation as%
\[
  \rho=\frac{\sigma_{\pi}^2}{\sigma_{\pi}^2+\sigma_{\varepsilon}^2},
\]%
it follows directly that we can alternatively parameterize effect size as%
\[
  \delta = \frac{\sqrt{\rho}\bigl(\max(\mu_j)-\min(\mu_j)\bigr)}{\sigma_{\pi}}.
\]%
Using this expression, I was able to set the marginal variance $\sigma^2_{\pi}+\sigma^2_{\varepsilon}$ to be constant across the varying values of our simulation parameters.  
    
\end{enumerate}

For each combination of number of observations ($n=20,50,80$), effect size ($\delta=0,0.2,0.5$), and intraclass correlation ($\rho=0.2,0.8$), I generated 1000 simulated datasets. For each of the datasets, I performed a repeated measures analysis of variance and, using the $F$ statistic and relevant values of $n$ and $k$, extracted two Bayes factors for $\mathcal{H}_0$; one based on the minimal BIC method of Equation $\ref{eq:bic2}$ and one based on the method of \citet{nathoo2016} which accounts for correlation between repeated measurements. These Bayes factors were then converted to posterior probabilities via Equation \ref{eq:probability}. To compare the performance of both methods in the simulation, I considered four analyses for each simulated dataset: (1) a visualization of the distribution of posterior probabilities $p(\mathcal{H}_0\mid \bm{y})$; (2) a calculation of the proportion of simulated trials for which the correct model was chosen (i.e., model choice accuracy); (3) a calculation of the proportion of simulated trials for which both methods chose the same model (i.e., model choice consistency); and (4) a calculation of the correlation between posterior probabilities from both methods.

\begin{figure}
  \centering
  \includegraphics[width=0.9\textwidth]{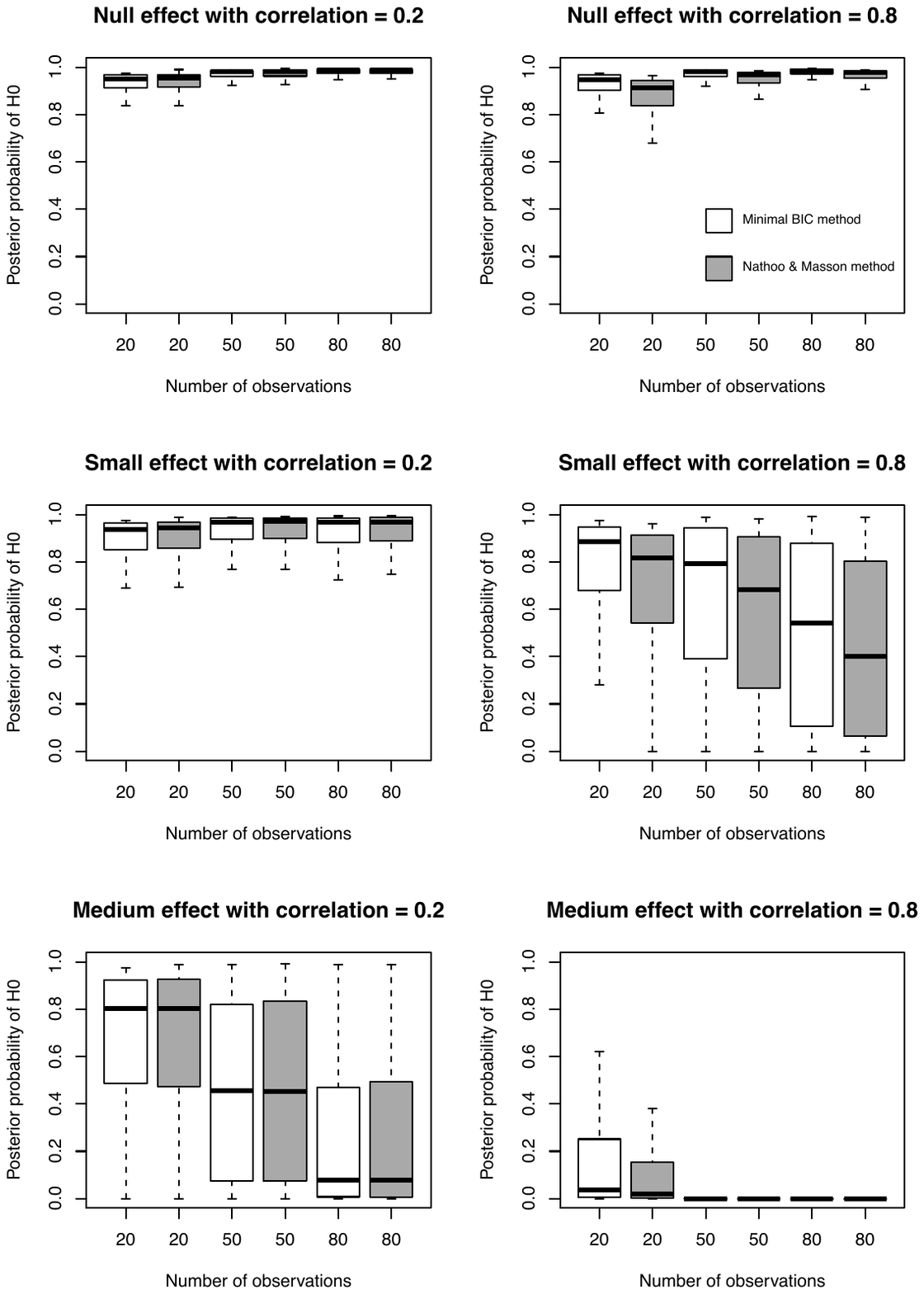}
  \caption{Results from our simulation. Each boxplot depicts the distribution of the posterior probability $p(\mathcal{H}_0\mid \bm{y})$ for 1000 Monte Carlo simulations. White boxes represent posterior probabilities derived from Bayes factors that were computed using the minimal BIC method of Equation \ref{eq:bic2}. Gray boxes represent posterior probabilities that come from the method of \citet{nathoo2016} which accounts for correlation between repeated measurements. }
  \label{fig:dist}
\end{figure}

First, let us visualize the distribution of posterior probabilities $p(\mathcal{H}_0\mid \bm{y})$. To this end, I constructed boxplots of the posterior probabilities, which can be seen in Figure \ref{fig:dist}. The primary message of Figure \ref{fig:dist} is clear. Our Equation \ref{eq:bic2}, which was derived from minimal BIC method developed in this paper appears to produce a distribution of posterior probabilities which is similar to those produced by the method of \citet{nathoo2016}. Moreover, this consistency extends across a variety of reasonably common empirical situations. In the cases where $\mathcal{H}_0$ was true (the first row of Figure \ref{fig:dist}, both Equation \ref{eq:bic2} and the \citet{nathoo2016} method produce posterior probabilities for $\mathcal{H}_0$ that are reasonably large. For both methods, the variation of these estimates decreases as the number of observations increases. When the intraclass correlation is small ($\rho=0.2$), the estimates from Equation \ref{eq:bic2} and the \citet{nathoo2016} method are virtually identical. When the intraclass correlation is large ($\rho=0.8$), the \citet{nathoo2016} method introduces slightly more variability in the posterior probability estimates. In all, these results indicate that Equation \ref{eq:bic2} is slightly more favorable when $\mathcal{H}_0$ is true.

For small effects (row 2 of Figure \ref{fig:dist}), the performance of both methods depended heavily on the correlation between repeated measurements. For small intraclass correlation ($\rho=0.2$), both methods were quite supportive of $\mathcal{H}_0$, even though $\mathcal{H}_1$ was the true model. This reflects the conservative nature of the BIC approximation \citep{wagenmakers2007}; since the unit information prior is uninformative and puts reasonable mass on a large range of possible effect sizes, the predictive updating value for any positive effect (i.e., $\text{BF}_{10}$) will be smaller than would be the case if the prior was more concentrated on smaller effects. As a result, the posterior probability for $\mathcal{H}_1$ is smaller as well. Regardless, the minimal BIC method (Equation \ref{eq:bic2}) and the \citet{nathoo2016} method produce a similar range of posterior probabilities. The picture is different when the intraclass correlation is large ($\rho=0.8$); both methods produce a wide range of posterior probabilities, though they are again highly comparable. It is worth pointing out that the posterior probability estimates all improve with increasing numbers of observations; but this should not be surprising, given that the BIC approximation underlying both the minimal BIC method and the \citet{nathoo2016} method is a large sample approximation technique. For medium effects (row 3 of Figure \ref{fig:dist}), we see much of the same message that we've already discussed previously. Both Equation \ref{eq:bic2} and the \citet{nathoo2016} method produce similar posterior probability values for $\mathcal{H}_0$. Both methods improve with increasing sample size, and at least for medium-size effects, the computations are quite reliable for high values of correlation between repeated measurements.

Though the distributions of posterior probabilities appear largely the same, it is not clear to what extent the two methods provide the user with an accurate inference. Since the data are simulated, it is possible to define a ``correct'' model in each case -- for simulated datasets where $\delta=0$, the correct model is $\mathcal{H}_0$, whereas when $\delta=0.2$ or $\delta=0.5$, the correct model is $\mathcal{H}_1$. To compare the performance of both methods, I calculated {\it model choice accuracy}, defined as the proportion of simulated datasets for which the correct model was chosen. Model choice was defined by considering $\mathcal{H}_0$ to be chosen whenever $\text{BF}_{01}>1$ and $\mathcal{H}_1$ to be chosen whenever $\text{BF}_{01}<1$. The results are displayed in Table \ref{tab:acc}.

\begin{table}
  \centering \small
  \begin{tabular}{ccccccc}
    & & \multicolumn{2}{c}{Correlation = 0.2} & & \multicolumn{2}{c}{Correlation = 0.8}\\
    & & Minimal BIC & Nathoo \& Masson & & Minimal BIC & Nathoo \& Masson\\
    \hline
    {\it Null effect}:\\
    $n=20$ & & .969 & .968 & & .979 & .954\\
    $n=50$ & & .989 & .988 & & .991 & .981\\
    $n=80$ & & .992 & .992 & & .992 & .985\\[2mm]

    {\it Small effect}:\\
    $n=20$ & & .068 & .072 & & .148 & .218\\
    $n=50$ & & .058 & .056 & & .307 & .374\\
    $n=80$ & & .062 & .062 & & .485 & .550\\[2mm]

    {\it Medium effect}:\\
    $n=20$ & & .259 & .266 & & .867 & .910\\
    $n=50$ & & .526 & .530 & & .997 & .999\\
    $n=80$ & & .760 & .756 & & 1.000 & 1.000\\
    \hline
    
  \end{tabular}
  \caption{Model choice accuracy for the minimal BIC method and the \citet{nathoo2016} method, calculated as the proportion of simulated datasets for which the correct model was chosen.}
  \label{tab:acc}
\end{table}

Let us consider Table \ref{tab:acc} in three sections. First, for data that were simulated from a null model, it is clear that the accuracy of both methods is excellent, with model choice accuracies all above 95\%. Further, the minimal BIC method outperforms the \citet{nathoo2016} method across all possible sample sizes as well as correlation conditions. However, the overall performance of both methods becomes more questionable for small effects. Model choice accuracies are no better than 5-7\% (regardless of sample size) for datasets with small correlation ($\rho=0.2$) between repeated measurements. The situation improves a bit when this correlation increases to 0.8, though never gets better than 55\%. Across all the small-effect datasets, the Nathoo and Masson method is slightly more accurate in choosing the correct model. This pattern continues for datasets which are simulated to have a large effect, though overall accuracy is much better in this case.

Overall, this pattern of results permits two conclusions. First, the BIC method (upon which both methods are based) tends to be conservative \citep{wagenmakers2007}, so the tendency to select the null model in the presence of small effects is unsurprising. Second, though performance was variable in the presence of small and medium effects, the differences in model choice accuracies between the minimal BIC method and the \citet{nathoo2016} method were small. Thus, any performance penalty that is exhibited for the minimal BIC method is shared by the Nathoo \& Masson method as well, reflecting not a limitation of the minimal BIC method, but a limitation of the BIC method in general. To further validate this claim, I calculated model choice {\it consistency}, defined as the proportion of simulated datasets for which both methods chose the {\it same} model. As can be seen in Table \ref{tab:con}, both the minimal BIC method and the Nathoo and Masson method choose the same model in a large proportion of the simulated datasets, regardless of effect size, sample size, or correlation between repeated measurements.

\begin{table}
  \centering \small

  \begin{tabular}{ccccc}
    & & Null effect & Small effect & Medium effect\\
    \hline
    {\it Correlation = 0.2}\\
    $n=20$ & & .997 & .994 & .977\\
    $n=50$ & & .999 & .994 & .984\\
    $n=80$ & & 1.000 & .998 & .994\\[2mm]

    {\it Correlation = 0.8}\\
    $n=20$ & & .975 & .930 & .957\\
    $n=50$ & & .990 & .933 & .998\\
    $n=80$ & & .993 & .935 & 1.000\\
    \hline

  \end{tabular}
  \caption{Model choice consistency for the minimal BIC method and the \citet{nathoo2016} method, calculated as the proportion of simulated datasets for which both methods chose the same model.}
  \label{tab:con}
\end{table}

\begin{figure}
  \centering
  \includegraphics[width=0.9\textwidth]{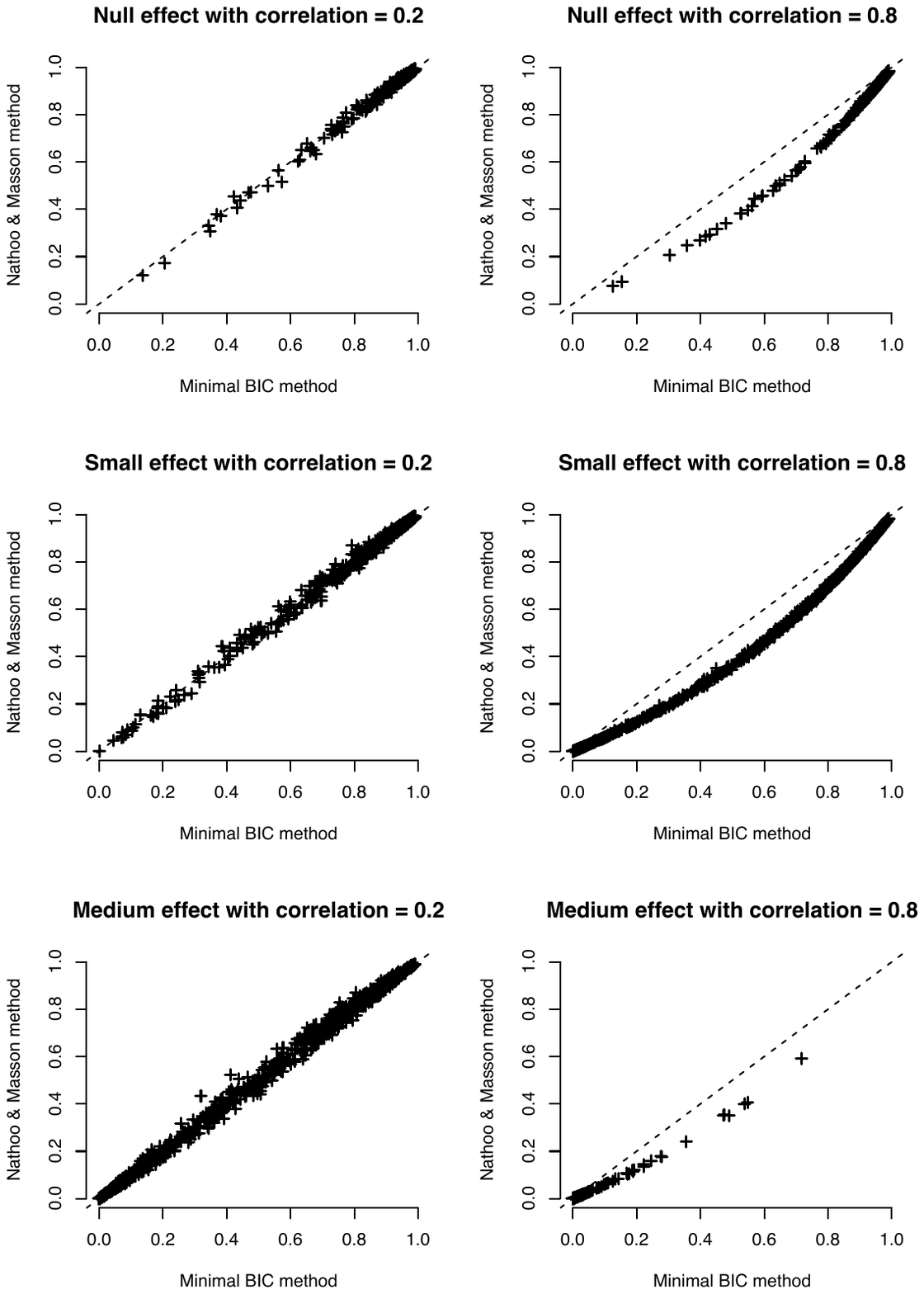}
  \caption{Scatterplot demonstrating the relationship between posterior probabilities calculated by the minimal BIC method (on the horizontal axis) and the \citet{nathoo2016} method (on the vertical axis). Sample size is assumed to be $n=50$ for all plots.}
  \label{fig:cors}
\end{figure}

As a final investigation, I calculated the correlations between the posterior probabilities that were produced by both methods. These correlations can be seen in Table \ref{tab:cors} and Figure \ref{fig:cors} -- note that the figure only shows scatterplots for the $n=50$ condition, though the $n=20$ and $n=80$ conditions produce similar plots. Table \ref{tab:cors} shows very high correlations between the posterior probability calculations. As can be seen in Figure \ref{tab:cors}, the relationship is linear when repeated measurements are assumed to have a small correlation, but nonlinear in the presence of highly correlated repeated measurements. For highly correlated measurements, the curvature of the scatterplot indicates that for a given simulated dataset, the posterior probability (for $\mathcal{H}_0$) calculated by the minimal BIC method will tend to be greater than the posterior probability calculated by the \citet{nathoo2016} method. Again, this is hardly surprising, as the Nathoo and Masson method is designed to better take into account the correlation between repeated measurements. One should note that this correction is advantageous for datasets generated from a positive-effects model, but disadvantageous for datasets generated from a null model. 

\begin{table}
  \centering \small

  \begin{tabular}{ccc}
    & Correlation = 0.2 & Correlation = 0.8\\
    \hline
    {\it Null effect}:\\
    $n=20$ & .993 & .987\\
    $n=50$ & .997 & .990\\
    $n=80$ & .998 & .988\\[2mm]

    {\it Small effect}:\\
    $n=20$ & .994 & .989\\
    $n=50$ & .998 & .991\\
    $n=80$ & .999 & .991\\[2mm]

    {\it Medium effect}:\\
    $n=20$ & .995 & .990\\
    $n=50$ & .999 & .995\\
    $n=80$ & .999 & .999\\
    \hline
  
  \end{tabular}
  \caption{Correlations between the posterior probabilities $p(\mathcal{H}_0\mid \bm{y})$ calculated by the minimal BIC method and the \citet{nathoo2016} method.}
  \label{tab:cors}
\end{table}

In all, the performance of the minimal BIC method is quite comparable to the \citet{nathoo2016} method. Though the Nathoo and Masson method is designed to better account for the correlation between repeated measurements, this advantage comes at a cost of increased complexity. On the other hand, the minimal BIC method introduced in this paper requires the user to only know the $F$-statistic, the number of subjects, and the number of repeated measures conditions. Thus, the small performance penalties for the minimal BIC method are far outweighed by its computational simplicity.

\section{Conclusion}

In this paper, I have proposed a formula for estimating Bayes factors from repeated measures ANOVA designs. These ideas extend previous work of \citet{faulkenberry2018}, who presented such formulas for between-subject designs. Such formulas are advantageous for researchers in a wide variety of empirical disciplines, as they provide an easy-to-use method for estimating Bayes factors from a minimal set of summary statistics. This gives the user a powerful index for estimating evidential value from a set of experiments, even in cases where the only data available are the summary statistics published in a paper. I think this provides a welcome addition to the collection of tools for doing Bayesian computation with summary statistics \citep[e.g.,][]{ly2018,faulkenberry2019}.

Further, I demonstrated that the minimal BIC method performs similarly to a more complex formula of \citet{nathoo2016}, who were able to explicitly estimate and account for the correlation between repeated measurements. Though the \citet{nathoo2016} approach is certainly more principled than a ``one-size-fits-all'' approach, it does require knowledge of the various sums-of-squares components from the repeated measures ANOVA, and though I have tried, I have not found an obvious way to recover the \citet{nathoo2016} estimates from the $F$ statistic alone. As such, the Nathoo and Masson approach is inaccessible without access to the raw data -- or at least the various SS components, which are rarely reported in empirical papers. Thus, given the similar performance compared to the \citet{nathoo2016} method, the new minimal BIC method stands at an advantage, not only for its computational simplicity, but also its power in producing maximal information given minimal input. 

\bibliography{references}
\bibliographystyle{apalike}

\end{document}